\documentstyle[12pt]{article}
\def\ie{\hbox{\it i.e.}{}}      \def\etc{\hbox{\it etc.}{}}
\def\eg{\hbox{\it e.g.}{}}

\font\elevenbf=cmbx10 scaled\magstep 1
 1
 1

\begin{document}
\newcommand{\cm}{Commun.\ Math.\ Phys.~}
\newcommand{\pr}{Phys.\ Rev.\ D~}
\newcommand{\pl}{Phys.\ Lett.\ B~}
\newcommand{\np}{Nucl.\ Phys.\ B~}
\begin{titlepage}
\begin{flushright}{\tt hep-ph/9510281}\\[5mm]
October 1995\end{flushright}
\vspace*{3.5cm}
\begin{center}
{\large\bf METHODS OF SUPERSYMMETRY BREAKING\\}
\vglue 5pt
\vskip 0.4in
{T.R. TAYLOR\\[1.5ex]}
{\it Department of Physics, Northeastern University\\}
{\it  Boston, MA 02115, U.S.A.\\}
\vglue 2cm
{\bf Abstract}
\end{center}
\vglue 0.3cm
{\rightskip=3pc
 \leftskip=3pc
 \noindent
This is a review of basic ideas and mechanisms encountered
in the supersymmetry breaking problem at the global level, in supergravity
models, and in superstring theory.\vglue 1cm}\vskip 1.5cm
\begin{center}
\it Invited talk presented at SUSY-95, Palaiseau, France, 15-19 May 1995
\end{center}
\end{titlepage}
\newpage
{\elevenbf\noindent 1. Introduction}
\vglue 0.2cm

This conference is a reunion of true SUSY believers,
so there is no need to argue that supersymmetry is really
a symmetry of particle physics. It is clear that we are all facing here
a long overdue problem why SUSY has not been seen at low energies.
If it is a ``good'', exact symmetry, it must be realised in a spontaneously
broken mode, because only in this case can we use it to make definite
predictions for superparticle masses and couplings. This is assuming
that we understand the origin of its breaking -- the super-Higgs mechanism.
Unfortunately, this part of the supersymmetric standard model
is still missing, which explains
the rather academic title of this talk; it is intended as an introduction
to the basic ideas and mechanisms of supersymmetry breaking.

In the standard model, electroweak symmetry is broken by a non-zero
vacuum expectation value (VEV) of the Higgs doublet.
In the case of supersymmetry, the analogous order parameters
are the VEVs of  auxiliary fields belonging
to either chiral or vector multiplets.
As explained in standard textbooks \cite{wb}, auxiliary fields
are introduced in order to close the off-shell supersymmetry algebra.
Under a supersymmetry transformation, the fermion $\psi$ belonging to a chiral
multiplet transforms as
\begin{equation}
\delta_{\epsilon}\psi=i\sigma^m\bar{\epsilon}\partial_mA+\epsilon F
\label{f}\end{equation}
where $A$ is the scalar partner of $\psi$ and $F$
is the auxiliary component of the multiplet. The latter does not
contain any physical degree of freedom. After using lagrangian
field equation, it becomes a function of physical fields:
$F=F(A,\dots)$. The VEV of $F$ is determined by further use
of equations of motion, including minimisation of the scalar potential \etc\
If it turns out to be non-zero,
\begin{equation}
\langle 0|F|0\rangle =F(\langle 0|A|0\rangle,\dots)\equiv M_{S}^2\
,\label{fvev}
\end{equation}
supersymmetry is broken spontaneously at mass scale $M_{S}$.
This is easy to understand. By looking at eq.(\ref{f}) we see that
\begin{equation}
\langle 0|[\epsilon Q,\psi ]|0\rangle   =
\langle 0|\epsilon F|0\rangle\ ,
\end{equation}
where $Q$ is the supercharge operator. Then
\begin{equation}
\langle 0|\epsilon F|0\rangle\neq 0~\Rightarrow~ \epsilon Q|0\rangle\neq 0
{}~\Rightarrow~ e^{i\epsilon Q}|0\rangle\neq|0\rangle\ , \label{arg}
\end{equation}
so the vacuum state which carries non-zero supercharge
is {\em not\/} invariant under supersymmetry transformations.
{}Futhermore, it can be shown that a massless fermion -- the goldstino --
must be present in the spectrum, populating
degenerate states obtained from the vacuum by SUSY transformations.
In the case of $F$-type breaking this is exactly the fermion $\psi$
which transforms under (\ref{f}) into the auxiliary field acquiring a
non-zero VEV.
Another type of SUSY breaking, the so-called $D$-type breaking may occur
in the presence of vector multiplets.
A vector multiplet contains a gauge boson and a gaugino $\lambda$ which
transforms as
\begin{equation}
\delta_{\epsilon}\lambda=\sigma^{mn}\epsilon F_{mn}+i\epsilon D\ ,\label{d}
\end{equation}
where $F_{mn}$ is the gauge field strength and $D$ is the auxiliary
component of the multiplet. By an argument similar to (\ref{arg}),
a non-zero VEV of $D$ breaks SUSY, with the gaugino identified as
the goldstino. If supersymmetry is gauged, \ie\ promoted to a local symmetry,
then the goldstino degrees of freedom are absorbed
by the massive spin $\textstyle\frac{3}{2}$ gravitino as
its helicity $\textstyle \pm \frac{1}{2}$ components.

The computation of $F$ and $D$ VEVs is a dynamical problem.
It may be simple in the case of weakly
interacting globally supersymmetric theories and supergravity, and possibly
more difficult in the presence of strong interactions, but the
basic idea is always the same: use field equations to determine
auxiliary VEVs. The form of field equations depends on a particular model.
The universal feature is the necessary presence of massless goldstinos
in spontaneously broken SUSY models.
This provides an intuitive criterion for SUSY breaking: the breaking
can occur only if there is a massless fermion in the spectrum --
a potential goldstino. The most sophisticated and rigorous
version of this argument is called the Witten index theorem \cite{wit}.
I will discuss separately the cases of global SUSY, supergravity
and superstring theory.
\vglue 0.4cm
{\elevenbf\noindent 2. Globally Supersymmetric Renormalisable QFT}
\vglue 0.2cm

A globally supersymmetric QFT is completely specified by the
superpotential $W(\Phi) $, an analytic function of chiral superfields
$\Phi$. The requirement of renormalisability restricts $W(\Phi)$ to a
polynomial of degree 3 in $\Phi$'s. The classical equations of motion for
the auxiliary fields are
\begin{equation}
\bar{F}_{\bar{\Phi}}=\frac{\partial W}{\partial\Phi}|_{\Phi=A}\ ,
\hskip 2cm  D_a=g_a\sum_{A}A^{\dag}T^aA+\xi^a\ ,\label{glob}
\end{equation}
where $g_a$ and $T^a$ are the gauge group couplings and generators,
respectively, and $\xi^a$ is the Fayet-Iliopoulos parameter
that may be non-zero for an index $a$ associated to a $U(1)$ subgroup only.
The classical scalar potential is
\begin{equation}
V(A)~=~\sum_{\Phi}|F_{\Phi}|^2+\sum_aD_{a}^2~~\geq 0\ ,\label{pot}
\end{equation}
with the auxiliary fields given by eq.(\ref{glob}). In the weak coupling
limit, the potential can be minimised to determine all VEVs
and to see whether supersymmetry is broken or not. For instance,
showing that $V>0$ in the vacuum is completely sufficient to prove
that some auxiliary fields acquire non-zero VEVs and hence SUSY is broken.
This procedure can be a posteriori justified if it happens that
all fields are weakly interacting at the SUSY breaking scale.
As usual, life is not so simple: it turns out that
supersymmetry remains unbroken in the minimal extension
of the supersymmetric standard model.
A completely new, ``hidden'' sector is necessary to trigger SUSY breaking.
{}For electroweak symmetry it was sufficient to
introduce one Higgs doublet with a simple potential, whereas
in the case of supersymmetry one needs at least several
chiral multiplets with a complicated superpotential and/or
{}Fayet-Illiopoulos terms associated with exotic $U(1)$'s,
each of them bringing in one more vector supermultiplet.
In these types of models, the supersymmetry breaking scale $M_S$
is introduced by hand. Another possibility is non-perturbative
supersymmetry breaking due to condensates, \ie\ non-zero VEVs
of composite fields playing the role of auxiliary components \cite{wit,dsb,a}.
$M_S$ can then be determined from the strong interaction scale
of ``supercolour'' forces that
cause dynamical supersymmetry breaking, which may seem to be
more natural than putting it in by hand. Supercolour
theories are not too difficult to construct; an important ingredient
is the absence of the mass gap, allowing the existence of
composite goldstinos.
The main problem, however, common to weakly and strongly coupled
hidden sectors, is how to communicate SUSY breaking to the observable sector
of quarks, squarks \etc\
A complicated system of ``messengers'' \cite{md} must be employed in order
to generate squark and gaugino masses.
The main virtue of this approach, advertised by its proponents,
is that the physics is fully contained below 1 TeV, staying away
from the traps and zasadzkas of quantum gravity, strings \etc\
In principle, this is a completely calculable scheme, but in practice
all viable models are very complicated and involve
a great deal of theoretical uncertainity.
\vglue 0.4cm
{\elevenbf\noindent 3. Local Supersymmetry and Standard Supergravity}
\vglue 0.2cm

As a consequence of the supersymmetry algebra which includes also the
momentum operator, gauging supersymmetry automatically brings
into the game gravity and the associated parameters --
the Planck mass $M_P\sim 10^{19}$ GeV and the coupling
$\kappa\sim 1/M_P$. The gravitino $\psi_{3/2}$
is the spin $\textstyle\frac{3}{2}$ gauge fermion of supersymmetry
which belongs the gravitational supermultiplet together
with the spin 2 graviton. All known forces can be unified
in the framework of supergravity. The theory is no longer renormalisable,
but as far as SUSY breaking is concerned, the lack of renormalisability
can be turned into advantage: higher dimensional interactions
provide a natural ``messenger'' system for communicating
SUSY breaking to the observable sector. Assuming that the scale
of SUSY breaking VEVs in the hidden sector is of order $\Lambda$,
and that higher-dimensional interactions
${\cal O}(\kappa^2)$ are responsible for the super-Higgs effects, we have
$m_{3/2}\sim\Lambda^3/M_{P}^2$. A gravitino mass of order of
1 TeV can be then generated by the hidden sector dynamics at
$\Lambda\sim 10^{13}$ GeV.

Together with relaxing the renormalisability requirement, there comes a
possible field-dependence of parameters
which are constrained to be constant in the global case.
It is encoded in the K\"ahler potential
$K$, in the superpotential $W$, and in the gauge functions $f_a$,
which depend on chiral superfields \cite{wb}. $W$ and $f_a$'s are analytic
while $K$ is real. The chiral superfields
(and the corresponding scalars), generically denoted by $A$, will be divided
into the observable ones - $q$, and the hidden ones - $\phi$.
In order to analyse SUSY breaking by hidden VEVs, it is convenient
to measure them in $M_P$ units, which can be done by a simple rescaling
that renders all $\phi$'s dimensionless.
The field dependence can be seen in the following formulas
for the wave-function factors ${\cal Z}$, Yukawa couplings
$Y$, and the gauge couplings $g_a$:
\begin{equation}
{\cal Z}_{I\bar{J}}(A)=\frac{\partial^2K(\phi, q)}{\partial
q^I\partial\bar{q}^{\bar{J}}}\ ,\hskip 1.5cm
Y_{IJK}(A)=\frac{\partial^3W(\phi, q)}{\partial q^I\partial q^J\partial q^K}\ ,
\hskip 1.5cm \frac{1}{g_{a}^2(A)}=\makebox{Re}f_a(\phi,q)\ .
\end{equation}
Since one is mostly interested in the VEVs of hidden fields,
expected to be much bigger than the observable VEVs,
one can expand in powers of $q$'s:
\begin{eqnarray}
K&=&\kappa^{-2}\widehat{K}(\phi)
+{\cal Z}_{I\bar{J}}(\phi)\, q^I\bar{q}^{\bar{J}}+\dots\ ,\\[3mm]
W&=&\widehat{W}(\phi)+ Y_{IJK}(\phi)\, q^Iq^Jq^K+\dots
\end{eqnarray}
Note that the hidden K\"ahler potential $\widehat{K}$ is
dimensionless while the superpotential $\widehat{W}$ has mass
dimension 3, therefore its size is set by the scale $\Lambda$
\ie\ $\widehat{W}\sim\Lambda^3$.
The supergravity version of the auxiliary field equations
(\ref{glob}) is
\begin{equation}
\bar{F}_{\bar{\phi}}=\kappa^2\, e^{\widehat{K}/2}
\left(\frac{\partial^2\widehat{K}}{\partial\phi\partial\bar{\phi}}\right)^{-1}
\left(\frac{\partial\widehat{W}}{\partial\phi}+
\widehat{W}\frac{\partial\widehat{K}}{\partial\phi}\right)+\dots\ \label{loc}
\end{equation}
The $D$ components are also given by expressions similar to
eq.(\ref{glob}), however since $F$-type breaking is very
easy to achieve, there is really no need to consider $D$-type breaking.

The formula for the scalar potential is slightly more complicated
than eq.(\ref{pot}), and there is one important difference:
it is not positive definite.
There is also another difference:
in order to find the vacuum it is no longer sufficient
to minimise this potential. The gravitational equations
of motion, which in the supergravity case play the role of gauge field
equations, are equally important. If the minimum of the potential
occurs at non-zero vacuum energy, the gravitational
background has a non-zero curvature. A flat Minkowski
background requires $V=0$ at the minimum,\footnote{Unless
one considers more complicated, cosmological solutions
with space-time dependent scalar fields.}
which unlike in the global case, turns out to be compatible with broken SUSY.
After ensuring that the classical minimum occurs at
$V=0$ at the classical level, it is not clear what to do with
quantum corrections. Because of this, the famous cosmological constant
inevitably gets in the way. There is no room for a separate
``adjusting'' of the cosmological constant without ruining
the mass relations \etc\ that follow from spontaneosly broken supersymmetry.
A possible procedure is to construct a model with a vanishing
tree-level cosmological constant, derive the spectrum, couplings \etc\,
and then analyse quantum corrections assuming the existence
of a physical ultraviolet cutoff \cite{mk,jb}.

Non-vanishing VEVs of hidden auxiliary components trigger spontaneous
supersymmetry breaking, generating the gravitino mass
\begin{equation}
m_{3/2}~=~ \left(\frac{1}{3}{\frac{\partial^2\widehat{K}}{\partial\phi
\partial\bar{\phi}}}\, F_{\phi}\bar{F}_{\bar{\phi}}
\right)^{1/2}=~ \kappa^2 \langle 0|
e^{\widehat{K}(\phi)/2}\,\widehat{W}(\phi)|0\rangle\ .
\end{equation}
It is important to be aware that the second part of this
equation, familiar to model-builders,
is correct under the assumption of $V=0$ at the minimum, it is
therefore  sensitive to a possible fine-tuning of
the cosmological constant. Note that a hidden superpotential
$\widehat{W}\sim\Lambda^3$ does indeed generate $m_{3/2}\sim\Lambda^3/M_P^2$.
As a result of higher-dimensional interactions between
the observable and hidden sectors implied by the
underlying supersymmetry, the observable scalars acquire masses ${\cal
O}(m_{3/2})$. The exact expressions for these masses depend on details
of the K\"ahler potential, \eg\ ${\cal Z}_{I\bar{J}}$ factors \etc,
therefore there is no reason to expect any special mass pattern.

When it comes to actual model building,
there is no problem with constructing SUSY-breaking hidden sector
superpotentials \cite{pran}. This can be achieved even
by one chiral multiplet with a linear superpotential, like in
the Polonyi model. In this case the hidden scale, hence effectively $M_S$,
is introduced by hand. A more ``natural'' scenario is offered by
no-scale models \cite{ns}, where the K\"ahler potential is adjusted
in such a way that a constant superpotential breaks supersymmetry
with an identically vanishing scalar potential at the tree level.
$M_S$ is determined then by radiative corrections to be of the same
order as the electroweak scale.
The bottom line is a softly-broken supersymmetric
low-energy effective field theory obtained from the
supergravity lagrangian by taking the limit $\kappa\rightarrow 0$
while keeping $m_{3/2}$ fixed \cite{ssb}. SUSY breaking can then
be parametrised by a finite number of parameters.
In the simplest supergravity models there are at least five such parameters:
universal scalar mass $m_0$, gaugino mass $m_{1/2}$,
higgsino mass parameter $\mu$, and two parameters, $A$ and $B$, which
specify the scalar potential. It is clear however that in
the absence of renormalisability there is no rigorous guiding principle
for selecting one hidden sector or another, therefore it is not
possible to make a definite prediction for the structure
of soft-breaking terms.

To summarise, supergravity provides a natural setting for SUSY breaking and
a messenger system for feeding this breaking down to
the supersymmetric standard model sector. On the other hand,
the lack of renormalisability and the cosmological constant
problem do clearly reduce its predictive power.
{}First of all, supergravity by itself gives no indication
about details of hidden sectors that are necessary to
derive the properties of low-energy
softly broken theory.  Furthermore, even if one starts
from a definite model at the classical level, it is not clear
whether a consistent treatment is possible for quantum corrections \cite{jb}.
Certainly, an ultraviolet cutoff is necessary in order to study the stability
of
the $M_P$ -- $M_S$ hierarchy and other phenomenological problems.
\vglue 0.4cm
{\elevenbf\noindent 4. Superstring Theory}
\vglue 0.2cm

There is only one or at worst a few superstring theories\footnote{
The reason for this hesitation should become clear at the end
of the talk.} --
heterotic, type I, II \etc\ -- but there are millions of
four-dimensional models corresponding to apparently degenerate
ground states of the same theory.
The present understanding of short-distance superstring dynamics
is not sufficient to select one particular model, or a class of
models, so it is better to pursue
a general analysis.
Each particular model contains one parameter, the
string mass scale $M\sim M_P$, and its low-energy limit
is described by a supergravity theory with definite
$K$, $W$ and $f$'s. The physical parameters like masses and couplings
depend then on the VEVs of hidden and observable scalar fields.
As far as SUSY breaking is concerned, the fundamental problem
is to understand how $M_S \ll M$ can be generated by these VEVs.
The breaking may involve effects associated with the extended string nature
or it may be simply a field-theoretical phenomenon.
\vglue 0.2cm
{\it\noindent 4.1. Stringy SUSY Breaking: Twisted and Magnetised Tori}
\vglue 0.2cm

Many four-dimensional superstring models can be constructed
by starting from ten dimensions and assuming that six
dimensions are compactified on a torus or another manifold.
The geometrical parameters that characterise compact dimensions are often
arbitrary.
In the effective field theory, this is reflected by the presence of massless
fields, the moduli, with the VEVs corresponding to six-dimensional
radii, angles \etc\ that remain undetermined at the classical level
due to flat directions of the scalar potential.
In addition to the massless modes, a typical spectrum also contains the towers
of Kaluza-Klein excitations with the masses quantised in units of inverse
radii $\sim 1/R$.

The simplest and in some sense unique mechanism for ``stringy''
SUSY breaking at an arbitrary scale is by twisting the compact tori, \ie\ by
imposing a special type of boundary conditions in compact dimensions
\cite{twist,rad}.
A typical twist cuts out every second state of each Kaluza-Klein tower and
eliminates the massless gravitino together with half of its tower.
The remaining half of the gravitino tower starts with a massive spin 3/2
particle which can be identified as the gravitino of spontaneously
broken supergravity with $m_{3/2}\sim 1/R$. In this way,
the SUSY breaking scale $M_S\sim 1/R$ becomes tied up with a compact
radius. From the supergravity point of view, twisted tori give
rise to stringy
realisations of no-scale models with vanishing potentials and zero cosmological
constant at the tree level.
SUSY breaking is due to a VEV of the auxiliary $F$-component of a
supermultiplet containing the modulus $T$ whose VEV determines $R$;
the modulino plays the role of the goldstino.
As mentioned before, $\langle 0|T|0\rangle=R$ remains arbitrary at the tree
level. This flatness of the potential is due to a special moduli-dependence
of the K\"ahler potential that follows directly
from superstring theory. Furthermore, the loop corrections have no
ultraviolet divergences since
the string mass scale $M$ provides a physical cutoff.

The usual pattern of soft-breaking terms induced by twisting is such that
the scalar partners of quarks and leptons remain massless
at the tree level whereas gauginos receive a common mass $m_{1/2}=m_{3/2}$
\cite{rad,kk}.
Once supersymmetry is broken,
the radiative corrections lift the flatness of the potential
by generating a non-trivial potential for $T$.
Minimisation of this potential with respect to $T$ and to the Higgs
field will fix their VEVs.
The new VEV scale is defined as the energy at which the mass squared
of the Higgs becomes negative and the breaking of electroweak
symmetry occurs, and is expected to be $\sim M_P\, e^{-1/Y^2(M_P)}$,
where $Y$ is some Yukawa coupling. In this way, $M_S\sim 1/R$
can be hierarchically smaller that $M_P$ provided that $h$ is not
too large.

A low supersymmetry breaking scale $M_S\sim 1$ TeV
corresponds to a large internal dimension. A completely
new, higher-dimensional world opens up above 1 TeV.
{} From the four-dimensional point of view, the extra dimensions
would manifest themselves by the presence of infinite towers
of Kaluza-Klein excitations. Na\"{\i}vely, this would seem to contradict
superstring unification at $10^{17}$ GeV which is based on the logarithmic
running of gauge coupling constants with the assumption of a desert
between $M_S$ and the unification mass. The reason why there is no
contradiction is that Kaluza-Klein states are organised in multiplets
of $N{=}4$ supersymmetry.
An $N{=}4$ multiplet contains one vector boson, four two-component spinors
and six real scalars. This leads to cancellation of the large radiative
corrections among particles of different spin and the evolution of gauge
couplings remains logarithmic, as in four-dimensional theory,
up to the Planck scale \cite{tv}.

The perspective of probing extra dimensions at future colliders seems
very appealing. Among the various Kaluza-Klein excitations of different
spin, the easiest to detect are the vectors
with the quantum numbers of the electroweak bosons.
They would decay into quarks, leptons or into their
SUSY partners; the lifetime can be estimated to be of order $10^{-26}$
seconds \cite{kk}.

There exists another way of SUSY breaking which employs extra dimensions.
A constant magnetic field, associated with a $U(1)$ gauge group,
which points in the direction of extra dimensions, generates mass splittings
within SUSY multiplets carrying non-zero $U(1)$ charges \cite{cb}.
Here again, $M_S\sim 1/R$. The main difference between
twisted and magnetised tori is that in the latter case
a non-zero potential, and a possible electroweak symmetry breaking,
are present already at the tree level.

To summarise, twisted and magnetised tori provide viable
mechanisms for low-energy SUSY breaking in superstring theory.
{}From the theoretical point of view the most important problem that requires
further clarification is the string description of the vacuum
rearrangement that leads to electroweak symmetry breaking
and to the determination of $M_S$. For instance, at the string
level, a non-vanishing one-loop cosmological constant leads to infinite
tadpoles at two loops, therefore a consistent prescription
for handling these divergences is necessary in order to
obtain definite predictions for the soft-breaking terms.
\vglue 0.2cm
{\it\noindent 4.2. Gaugino Condensation}
\vglue 0.2cm

{}Four-dimensional superstring theories usually contain
very rich spectra that include not only the standard
model sector but also hidden sectors which are very often
associated with a whole new non-abelian gauge group.
Dynamical supersymmetry breaking may then occur
as a non-perturbative effect of hidden gauge interactions,
much like in the supercolour idea mentioned before, and may be communicated
to the observable sector via higher-dimensional interactions.
Assuming that non-perturbative
effects take place at energies much lower than $M\sim M_P$,
they can be described within the framework of the effective
field theory. This approach can only be justified {\it a posteriori\/}:
once supersymmetry is found broken at $M_S\ll M$,
one should argue that the respective physical mechanism
remains unaffected by high-energy string physics.

As an example of a simplest hidden gauge sector, consider
an asymptotically free QFT defined by a pure
supersymmetric Yang-Mills system with an arbitrary gauge group.
Non-perturbative dynamics of this theory have been studied extensively
in the past in the context of global supersymmetry.
In particular, there is a mass gap, and the lightest
fermion, which is expected to be the superpartner of the glueball,
has a mass of order of the strong interaction scale
$\Lambda$ \cite{vy}. Since there is no goldstino available, supersymmetry
remains unbroken even at the non-perturbative level,
as confirmed by Witten index theorem \cite{wit}. On the other hand,
a non-perturbative effect that does certainly occur is
the gaugino condensation \cite{vy} which gives rise to
\begin{equation}
\langle 0|\lambda\lambda|0\rangle~\sim~\Lambda^3 ~\sim~
\mu^3\, \exp({\textstyle -\frac{3}{2\beta_0\, g^2(\mu)}}) \label{cond}
\end{equation}
where $\mu$ is the renormalisation scale, $g(\mu)$ is the gauge coupling,
and $\beta_0$ is the one-loop beta function coefficient.

There is an important difference between globally supersymmetric
gauge theories and the effective field theories describing
gauge interactions in superstring theory. In the latter case,
the gauge couplings, similarly to other physical quantities,
correspond to dynamical parameters which are determined by
VEVs of some scalar fields.
In heterotic superstring theory, a typical gauge function which determines
the gauge coupling at the string scale has the form
\begin{equation}
f_a(\phi) ~=~ S+f_{a}^{(1)}(T) ~=~ \frac{1}{g_{a}^2(S,T)}\ ,\label{coup}
\end{equation}
where the tree-level contribution depends universally on the
dilaton $S$ while the one-loop threshold corrections $f_{a}^{(1)}$
depend on the moduli $T$ \cite{dkl,ant}.
As a result, the auxiliary field equations
receive additional terms involving gaugino bilinears:
\begin{equation}
\bar{F}_{\bar{\phi}}~=~ \bar{F}_{\bar{\phi}}
(\makebox{\small{BOSONS}})+\kappa^2\,
\left(\frac{\partial^2\widehat{K}}{\partial\phi\partial\bar{\phi}}\right)^{-1}
\frac{\partial ({\textstyle\frac{1}{g^2(\phi)}})}{\partial\phi}\,
\langle 0|\lambda\lambda|0\rangle+\dots\ ,    \label{auxi}
\end{equation}
where $F_{\phi}(\makebox{\small{BOSONS}})$ is the bosonic part given by
eq.(\ref{loc}). In this way, gaugino condensation breaks supersymmetry
at $M_S\sim \Lambda^3/M_P^2$ \cite{gc}.
The missing goldstino is found as a combination of the dilatino
and the modulinos, as seen from eqs.(\ref{coup}) and (\ref{auxi}).

The values of gauge couplings at the string scale, hence
$\Lambda$ and $M_S$, are all determined by the dilaton
and moduli VEVs. In order to compute these VEVs one has to determine
first the effective potential induced by non-perturbative
effects. This can be done by integrating out
the gauge degrees of freedom in the effective theory describing a coupled
Yang-Mills -- dialton/moduli system \cite{tay}.
The final result is the effective
superpotential
\begin{equation}
\widehat{W}(S,T)~\sim~\Lambda^3 ~\sim~  M^3\,
\exp({\textstyle -\frac{3}{2\beta_0\, g^2(S,T)}})\ .\label{sup}
\end{equation}
The moduli-dependence of $\widehat{W}$ and of the respective
potential is due to the one-loop threshold corrections $f_{a}^{(1)}(T)$,
eq.(\ref{coup}). The form of these functions is well known,
however there is no need to go into details to point out some basic
features of the potential.
The strongest constraint comes from the invariance
of superstring theory under duality transformations $R\rightarrow
1/(RM^2)$ relating large and small radius compactifications.
This duality is due to a complete symmetry between
Kaluza-Klein excitations and string winding modes. It is reflected in
the effective field theory, hence in the scalar potential,
as a symmetry under modular transformations  $T\rightarrow 1/T$.
A potential symmetric under such a transformation has an obvious
stationary point at the self-dual point $T=1$. A more detailed analysis,
using the explicit expressions for threshold corrections, shows
that this corresponds to a minimum or that a true minimum
with respect to $T$ is located in the neighbourhood of the
self-dual point. In this way, the radii are stabilised at
a typical value $R\sim 1/M$.

On the other hand, the minimisation of the potential with respect to
the dilaton $S$ presents a more difficult problem.
{}From the dilaton-dependence of gauge couplings, eq.(\ref{coup}),
it follows that $\widehat{W}\sim\exp(-3S/2\beta_0)$.
The respective potential falls off exponentially at large $S$
and there is no stable minimum. There is of course a ``runaway''
vacuum at $S\rightarrow\infty$ corresponding to $\Lambda\rightarrow 0$
and unbroken supersymmetry. It is not surprising
that the theory prefers to relax in a zero energy supersymmetric vacuum.
It is very difficult to understand how a stable vacuum can
exist at finite $S$. The formula (\ref{coup})
which is responsible for the exponential suppression
of the superpotential is correct to all orders of perturbation theory
\cite{ant}.
A different dilaton dependence of gauge couplings, induced by some truly
non-perturbative superstring effects, could in principle
alter eq.(\ref{coup}) \cite{filq,sd}. However, a low $M_S$ requires
gaugino condensation to occur at $\Lambda\ll M$,
and it is hard to imagine how genuinely superstring effects could interfere
at such a low scale. The onset of these efects can be
seen in the effective field theory as the appearance of interactions
described by higher-derivative supergravity,
but all of them are suppressed by the powers of $\Lambda/M$.

To summarise, there is a serious self-consistency
problem with QFT description of SUSY breaking by
gaugino condensation.\footnote{This problem may be absent though in some
models,
with gauge groups consisting of several non-abelian factors \etc\ }
There is much further work needed in order to provide superstring-theoretical
description of non-perturbative QFT physics.
{}From this point of view, recent developments in dualities and other
non-perturbative aspects of superstring theory
look very promising and go straight in the right direction.
\vglue 0.4cm
{\elevenbf\noindent 5. New Results and Perspectives}
\vglue 0.2cm

Up to this point,
there have not been many new results reported in this review.
In the past year, most of field-theoretical studies of SUSY
breaking have focussed on the following topics:
\begin{itemize}
\setlength{\itemsep}{-.9pt}
\item model-building with dynamical SUSY breaking \cite{md}
\item general analysis of soft-breaking terms in the effective
supergravity theory \cite{sb}
\item studies of the effective actions describing gaugino condensation
\cite{efa,bgt}
\item mass generation for the universal axion \cite{bgt}
\item strong-weak coupling duality-inspired dilaton stabilisation \cite{sd}.
\end{itemize}

Recently, there have been many exciting new developements in superstring
theory that bear excellent prognosis for a deeper understanding of SUSY
breaking. Many mysterious ``dualities'' \cite{d} have been discovered
which allow exact determination of some physical quantities
in $N=2$ and $N=1$ supersymmetric models. For instance,
a $N=2$ prepotential which usually contains perturbative and non-perturbative
contributions can be computed in some cases exactly as a purely classical
quantity in the dual theory \cite{kv}.
All dualities known so far relate theories
with equal  number of supersymmetries, so they are not useful
for SUSY breaking. There is no reason however why
dual descriptions should not exist for $N=1$ superstrings with
Yang-Mills sectors that break supersymmetry by gaugino condensation.
It would not be surprising if the dual descriptions involved twisted or
magnetised tori; the two previous subsections might in fact describe different
aspects of the same physical mechanism.

In summary, there is a clear advantage gained by promoting
supersymmetry to a local symmetry: all known intractions
can be described in one unified framework of supergravity.
In supergravity models, SUSY breaking is transmitted from the hidden sector
to the observable sector in a very natural way.
Superstrings take us much farther, by offering
a completely calculable framework with a physical short-distance cutoff. Many
important aspects of SUSY breaking in superstring theory have already been
understood.
It remains however to put several pieces together to obtain
a fully consistent picture; most likely, it will include
some sort of superstring -- nonsupersymmetric string dualities.
In this way, superstring theory may finally offer some
firm predictions that can be tested at future colliders.
\vglue 0.4cm
{\elevenbf\noindent Acknowledgements}
\vglue 0.2cm

I thank the organisers of this conference for their hospitality
and for providing a most stimulating atmosphere. This work was supported
in part by the National Science Foundation under grant PHY-93-06906.

\end{document}